\documentclass[8pt]{article}
\usepackage{times}

\usepackage{amssymb}

\usepackage[figuresright]{rotating}

\begin{document}





\begin{center}
{\Large \bf Nuclear PDFs at NLO -- status report and review of the EPS09 results}
\vspace{0.5cm}


{\large K. J. Eskola$^{a,b,}\footnote{Speaker at the \textit{Hard Probes 2010}, Eilat, Israel, October 2010.}$, H. Paukkunen$^c$ and C.A. Salgado$^c$}\\
\vspace{0.5cm}

{\small 
$^a$Department of Physics, P.O. Box 35, FI-40014 University of Jyv\"askyl\"a, Finland\\
$^b$Helsinki Institute of Physics, P.O. Box 64, FI-00014 University of Helsinki, Finland\\
$^c$Departamento de F\'\i sica de Part\'\i culas and IGFAE, Universidade de Santiago de Compostela, Spain\\
}
\vspace{0.5cm}

\end{center}


\begin{abstract}
We review the current status of the global DGLAP analysis of nuclear parton distribution functions, nPDFs, focusing on the recent EPS09 analysis \cite{EPS09}, whose output, EPS09NLO, is the best-constrained NLO nPDF set on the market. Collinear factorization is found to work very well in the kinematical region studied. With the error sets released in the EPS09 package one can compute how the nPDF-related uncertainties propagate into factorizable nuclear hard-process cross sections. A comparison with the other existing NLO nPDF sets is shown, and the BRAHMS forward-$\eta$ hadron data from d+Au collisions are discussed in the light of the EPS09 nPDFs and their error sets.
\end{abstract}







\section{Introduction}
\label{sec:intro}

In this talk, I will discuss the nPDFs which are obtained through genuinely global analyses, which are based on the DGLAP evolution \cite{DGLAP} and where one uses various types of nuclear hard process data in extracting the nPDFs. Such analyses are a test of perturbative QCD (pQCD) and collinear factorization, schematically expressed as
\begin{equation}
\sigma_{AB\rightarrow h+X} = \sum_{i,j} f_i^A(Q^2)\otimes f_j^B(Q^2)\otimes \sigma_{ij\rightarrow h+X'}(Q^2)+ {\cal O}(1/Q^2),
\label{eq:factorization}
\end{equation}
where the factorization/renormalization scales are sufficiently large, $Q^2\gg \Lambda_{\rm QCD}^2$, and the PDFs $f_i^A, f_j^B$ are universal and transferable from one hard process to another. 

Traditionally, both for the free and bound proton PDFs, the global DGLAP analysis procedure has been as follows: one starts with the non-perturbative input, an initial assumption, for the PDFs $f_i(x,Q_0^2,\{a_i\})$, which are expressed in terms of a set of fit parameters $\{a_i\}$ and where the initial scale $Q_0$ typically is of the order of 1 GeV. QCD sum rules are imposed at this stage. The PDFs at larger scales, $f_i(x,Q^2>Q_0^2,\{a_i\})$, are obtained through the standard pQCD DGLAP evolution to the required order (LO, NLO,...). After this, one compares the various hard-process cross sections, which are computed using the set $\{f_i(x,Q^2,\{a_i\})\}$, with existing data. The best set of PDFs is determined through iteration in the $a_i$-parameter space, by finding the set $\{a_i\}$ which minimizes the $\chi^2$ of the fit. After this, one has to perform a rather non-trivial error analysis to determine the non-correlated "error PDF sets" which allow one to study the propagation of the nPDF-uncertainties into hard cross sections. Thus, the outcome of such global analysis is the best PDF set $\{f_i(x,Q^2\ge Q_0^2)\}$, supplemented with a number of error sets.  

Here I will not be addressing the QCD origin of the nuclear effects in PDFs -- for discussion on this important question see e.g. Refs.~\cite{Armesto:2006ph,Frankfurt:2003zd,Tywoniuk:2007xy,Armesto:2010kr}. I will focus on the linear DGLAP evolution only, studies of other evolution equations (BFKL, BK,...), saturation or nonlinear effects \cite{Gribov:1984tu} will not be addressed here either. As indicated by Eq.~(\ref{eq:factorization}),  power corrections (see e.g. Refs.~\cite{Qiu:2003vd,Kulagin:2004ie}) will be ignored also. 

The global analyses of PDFs and nPDFs are quite challenging, since in general the data lie in correlated regions of $x$ and $Q^2$, and since there are statistical, systematic and also additional normalization errors to deal with but no unique way of propagating these uncertainties into the PDFs. As the parameter space is 15--30 dimensional, \textit{very} fast DGLAP and cross-section solvers are needed especially in the NLO (for technical details, see \cite{Paukkunen:2009ks}). 

The global DGLAP analyses have resulted in excellent fits for collinearly factorized free proton PDFs: sets like CT10 \cite{Lai:2010vv}, MSTW \cite{Martin:2009iq}, and (from neural networks) NNPDF2.0 \cite{Ball:2010de} are nowadays available. The global analysis of nPDFs, however, is even more challenging than that of the free proton PDFs, since in addition to the $x$ and $Q^2$ dependence one needs to consider also the $A$ dependence of the PDFs. There also is a smaller amount of data from fewer types of processes available as constraints, and the kinematical coverage of the nuclear hard-process data is more limited than in the free proton PDF case. 

Genuinely global analyses of nPDFs have been performed now over a decade. The pioneering LO analysis, where the $\chi^2$ minimization was made by eye, resulted in the nPDF set EKS98 \cite{EKS98}, which still is valid and useful. The first error estimates on nPDFs were provided for the LO sets HKM \cite{Hirai:2001np} and HKN04 \cite{Hirai:2004wq}. The first NLO nPDFs were given in the nDS set \cite{deFlorian:2003qf} and the first error estimates in the NLO case in the HKN07 set \cite{Hirai:2007sx}. The latest progress was made in the EPS09 analysis \cite{EPS09} which includes RHIC data and
now for the first time provides, in addition to the best NLO fit, 30 concrete nPDF error sets (for each $A$). With EPS09, which currently represent the state-of-the-art NLO nPDFs, one has reached the same degree of technical rigor in the global analysis for nPDFs as achieved already some time ago in the free proton case.

\section{EPS09 global analysis framework}
\label{sec:EPS09_framework}

Let me next describe the main features in the EPS09 analysis framework,  for details, see the Ref.~\cite{EPS09}. We define the bound proton PDFs relative to the free proton ones, taking the set CTEQ6.1M \cite{Stump:2003yu} as our baseline:
\begin{equation}
f_{i}^A(x,Q^2) \equiv R_{i}^A(x,Q^2) f_{i}^{\rm CTEQ6.1M}(x,Q^2),
\end{equation}  
where $i$ is the parton flavour. Choosing CTEQ6.1M means that we work in the $\overline{MS}$ and zero-mass variable flavor-number schemes.  Isospin symmetry is assumed, so that the average $u$-quark distribution in a nucleus $A$ becomes $u_A(x,Q^2) = \frac{Z}{A} f_u^A(x,Q^2) + \frac{A-Z}{A} f_d^A(x,Q^2)$, and similarly for the $d$-quark and $\bar u,\bar d$ antiquarks. 

As illustrated in Fig.~1 (left), one has to deal with the following $x$-, $Q^2$- and $A$-dependent nuclear effects in the nPDFs: the "shadowing" depletion observed in the DIS structure function ratios $\frac{1}{A}F_2^A/\frac{1}{2}F_2^D$ at $x\ll 0.1$, the "antishadowing" excess around $x\sim 0.1$, the "EMC-effect" depletion at $0.2<x<0.7$ and the "Fermi motion" excess towards $x=1$ and beyond. These effects are in the nPDFs at our (CTEQ6.1M's) initial scale $Q^2_0=1.69$~GeV$^2$.  In the lack of sufficient data constraints, we are forced to start with only three different modification ratios: $R_G^A(x,Q_0^2)$, $R_V^A(x,Q_0^2)$, and $R_S(x,Q_0^2)$ for gluons, valence quarks and sea quarks, correspondingly. The $A$ dependence is embedded in the $A$ dependence of the parameters, such as the one controlling the  antishadowing: $y_a=y_a({\rm C})(A/12)^{p_a}$, with carbon as the reference nucleus. The small nuclear effects in deuterium are neglected.

In the EPS09 analysis, we have altogether 929 data points from 32 different data sets at our disposal. Three types of data are used: deep inelastic lepton-nucleus scattering (DIS), Drell-Yan (DY) dimuon production in p+$A$ collisions from FNAL, and most lately -- and included so far only in EPS08 (LO) \cite{EPS08} and in EPS09 (NLO, LO) \cite{EPS09} -- also neutral-pion production at mid-rapidity in $d$+Au collisions from BNL-RHIC. Note that unlike in EPS08, we do not utilize the BRAHMS forward-rapidity hadron data from d+Au in the EPS09 analysis. 

The goodness measure $\chi^2$, which is minimized as described in Sec.~\ref{sec:intro}, is here a generalized one \cite{Stump:2001gu,EPS08},
\begin{equation}
\chi^2(\{a\})    \equiv  \sum _N w_N \, \chi^2_N(\{a\}), \quad {\rm where}\quad
\chi^2_N(\{a\})  \equiv  \left( \frac{1-f_N(\{a\})}{\sigma_N^{\rm norm}} \right)^2 + \sum_{i \in N}
\left[\frac{ f_N(\{a\}) D_i - T_i(\{a\})}{\sigma_i}\right]^2,
\label{eq:chi2}
\end{equation}
where we give an additional weight $w_N$ to those data sets $N$ which provide important constraints  but whose number of data points is so small that these constraints would otherwise escape unnoticed.  This standard weighting procedure does not cause a bias as long as no significant tension between different data sets arises. 
The multiplier $f_N(\{a\})$, which minimizes the $\chi^2_N(\{a\})$ for a data set $\{D_i\}_N$ at every round of iteration (needed only for the RHIC data now), is an output of the analysis, giving the best estimate of the overall normalization which is still consistent with the error $\sigma_N^{\rm norm}$ provided by the experiment. 

\begin{figure}[h!]
\vspace{-0.3cm}
\center
\includegraphics[scale=0.6]{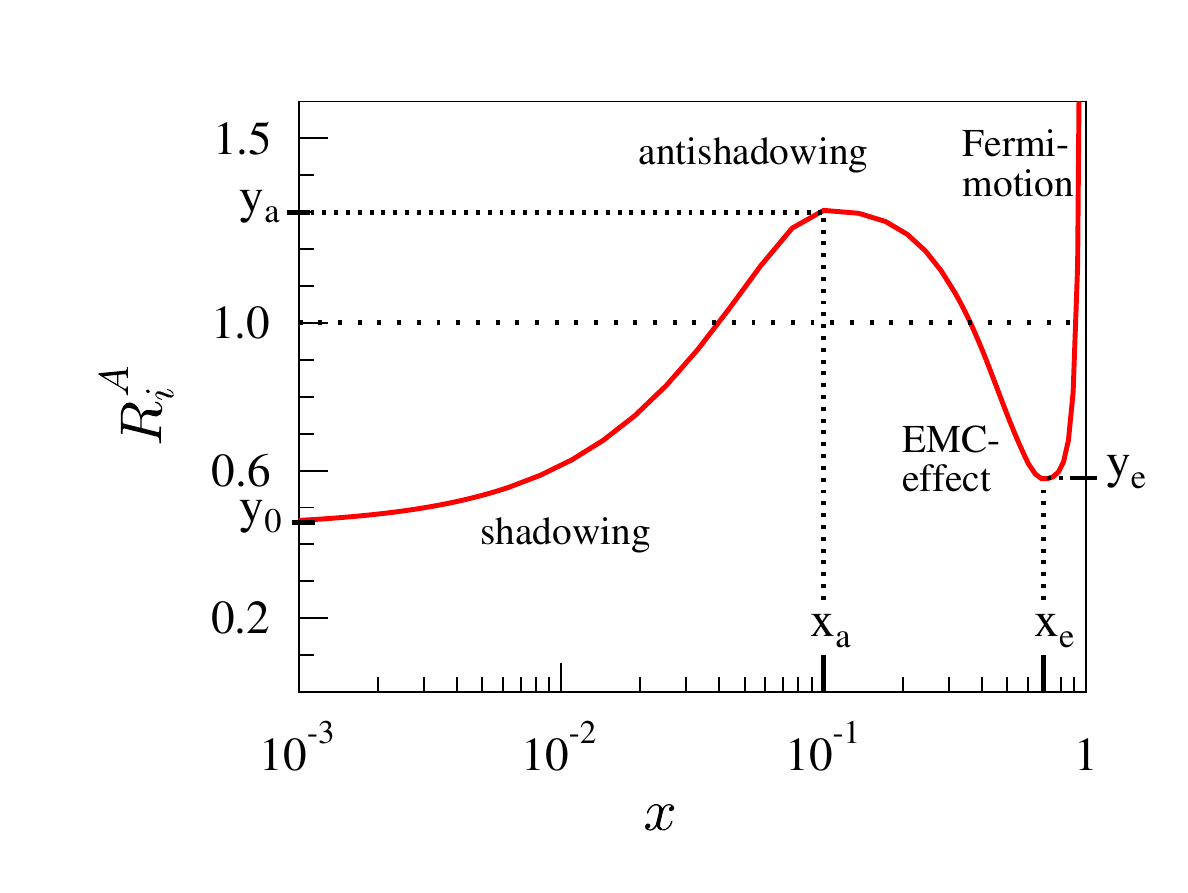} 
\includegraphics[scale=0.35]{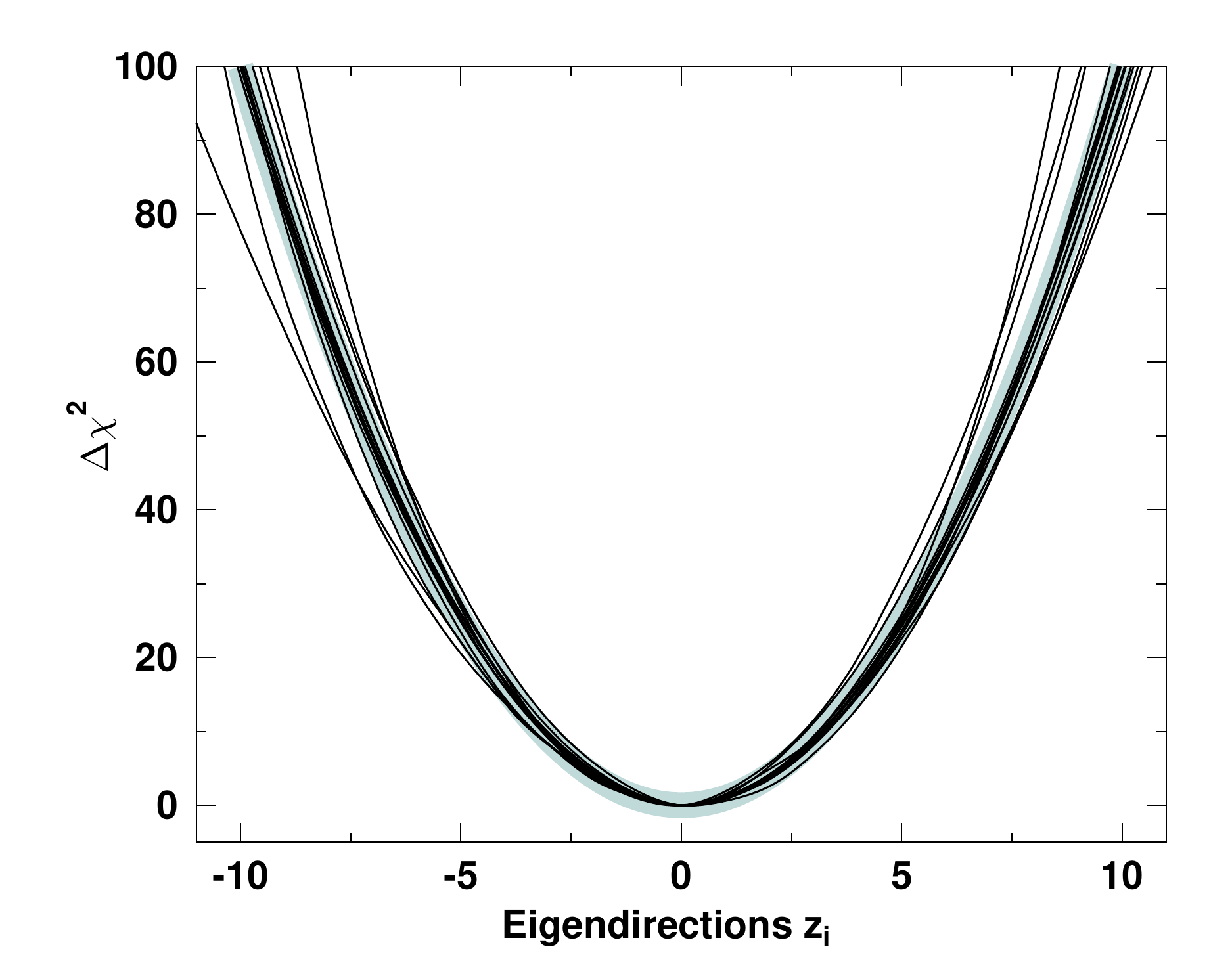} 
\includegraphics[scale=0.30]{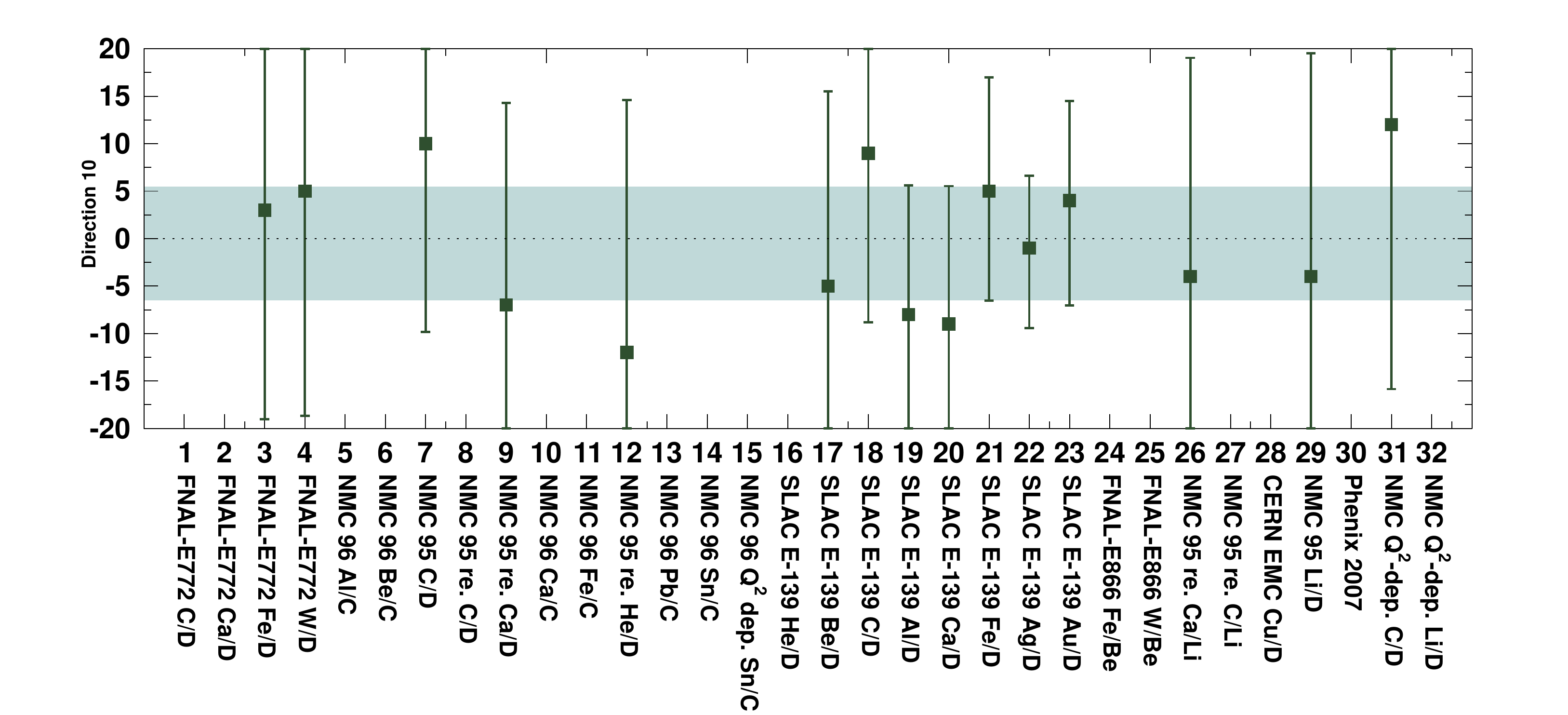}
\vspace{-0.2cm}
\caption[]{Elements in the EPS09 analysis.
\textbf{Left:} A sketch of the EPS09 fit functions $R_i^A(x)$ and the role of certain fit parameters. 
\textbf{Right:} $\chi^2-\chi^2_0$ as a function of each linear combination $z_i$, and its comparison with the ideal quadratic behaviour $\Delta \chi^2 = z_i^2$ (the shaded band). 
\textbf{Bottom:} The 90\% confidence limits for each data set in the eigendirection $z_{10}$. The filled boxes show the location of the $\chi^2$ minimum for each data set, while the global minimum is at $z=0$. 
All panels are from \cite{EPS09}.}
\label{fig:chi2}
\end{figure}

We minimize the global $\chi^2$ with respect to 15 fit parameters $a_i$, which quite obviously are correlated with each other. The minimum is found to be at $\chi^2_0=731.3$ for 929 data points. Once $\chi^2_0$ (the best fit) is determined, we perform an error analysis using the Hessian method. First, to find those linear combinations of $a_i$ which are uncorrelated, we diagonalize the Hessian matrix \textbf{H}, whose elements are the second derivatives of $\chi^2$ at $\chi^2_0$. Then, denoting by $z_i$ the suitably normalized parameter eigendirections, we can expand the $\chi^2$ as follows:
\begin{equation}
\chi^2 
= \chi_0^2 + \sum_{ij} \frac{1}{2} \frac{\partial^2 \chi^2}{\partial a_i \partial a_j}\Big|_{a=a^0}\delta a_i\delta a_j + {\cal O}(\{\delta a_i\}^3)
\approx \chi_0^2 + \sum_{ij}\delta a_i H_{ij} \delta a_j = \chi_0^2 + \sum_{i} z_i^2,
\label{eq:QuadraticApprox}
\end{equation}
After this, the error propagation becomes more straightforward and we can obtain the error for a physical nPDF-dependent 
quantity $X$ through the variations of the uncorrelated parameters $z_i$ as (for details, see \cite{EPS09}), 
\begin{equation}
(\Delta X)^2 \approx \sum_j \left( \frac{\partial X}{\partial z_j} \right)^2 (\delta z_j)^2 = 
\Delta \chi^2 \sum_j \left( \frac{\partial X}{\partial z_j} \right)^2,
\end{equation}
where we have taken the same increase, $\Delta \chi^2=(\delta z_i)^2$, in each eigendirection $i$ of \textbf{H}. Figure \ref{fig:chi2} (right) shows that the quadratic approximation in Eq.~(\ref{eq:QuadraticApprox}) above is working well sufficiently near the $\chi^2$ minimum. No unique procedure, however, exists for obtaining $\Delta \chi^2$. We use the "90\% confidence criterion" originally developed by CTEQ \cite{Pumplin:2002vw}, and explained in detail in \cite{EPS09}. The idea, illustrated in Fig.~\ref{fig:chi2} (bottom), is as follows: the $\chi^2$ minimum for each data set of $N$ data points is first determined separately for each eigendirection $i$. Then, we let the parameter $z_i$ grow(decrease) from the extremum location up(down) to the point that the $\chi^2$ of this data sets reaches the value below which the $\chi^2$ should reside within a 90\% probability, if an $N$-dimensional Gaussian probability distribution for the $\chi^2$ is assumed. Repeating this procedure for all the data sets, and in the end taking the stringest of all the obtained upper/lower limits, $(\delta z_i^{\pm})^2 = \Delta\chi^2(\delta z_i^{\pm})$ for each data set, and averaging over all eigendirections $i$, we arrive at \begin{equation}
 \Delta \chi^2 \equiv \sum_i \frac{\Delta \chi^2 (\delta z_i^+) + \Delta \chi^2 (\delta z_i^-)}{2N} \approx \sum_i \frac{(\delta z_i^+)^2 + (\delta z_i^-)^2}{2N} \approx 50.
\label{eq:finalDeltachi2}
\end{equation}
Whether $\Delta\chi^2=50$ above reproduces the average size of the actual error bars in the data, needs of course be checked a posteriori. From Fig.~\ref{fig:chi2} (right), however, we see that for this choice of $\Delta\chi^2$, the quadratic approximation is still fine.

Thus, the EPS09 nPDF package contains the central set (best fit) $S^0$ and 15+15 error sets $S_i^{\pm}$, which are obtained by changing the fit parameters along the plus and minus directions of $z_i$ so that the global $\chi^2$ grows by 50. Since the error sets $S_i^{\pm}$ may shift a physical quantity $X$ which the user wishes to study, into the same direction, we recommend the following prescription \cite{Nadolsky:2001yg} for computing the upper and lower limits of $X$: 
\begin{eqnarray}
(\Delta X^+)^2  & \approx &  \sum_i \left[ \max\left\{ X(S^+_i)-X(S_0), X(S^-_i)-X(S_0),0 \right\} \right]^2,\nonumber\\ 
(\Delta X^-)^2  & \approx & \sum_i \left[ \max\left\{ X(S^0)-X(S^+_i), X(S^0)-X(S^-_i),0 \right\} \right]^2.
\label{eq:X_Extremum3}
\end{eqnarray}

\begin{figure}[hbt]
\center
\vspace{-.5cm}
\hspace{-.7cm}
\includegraphics[scale=0.45]{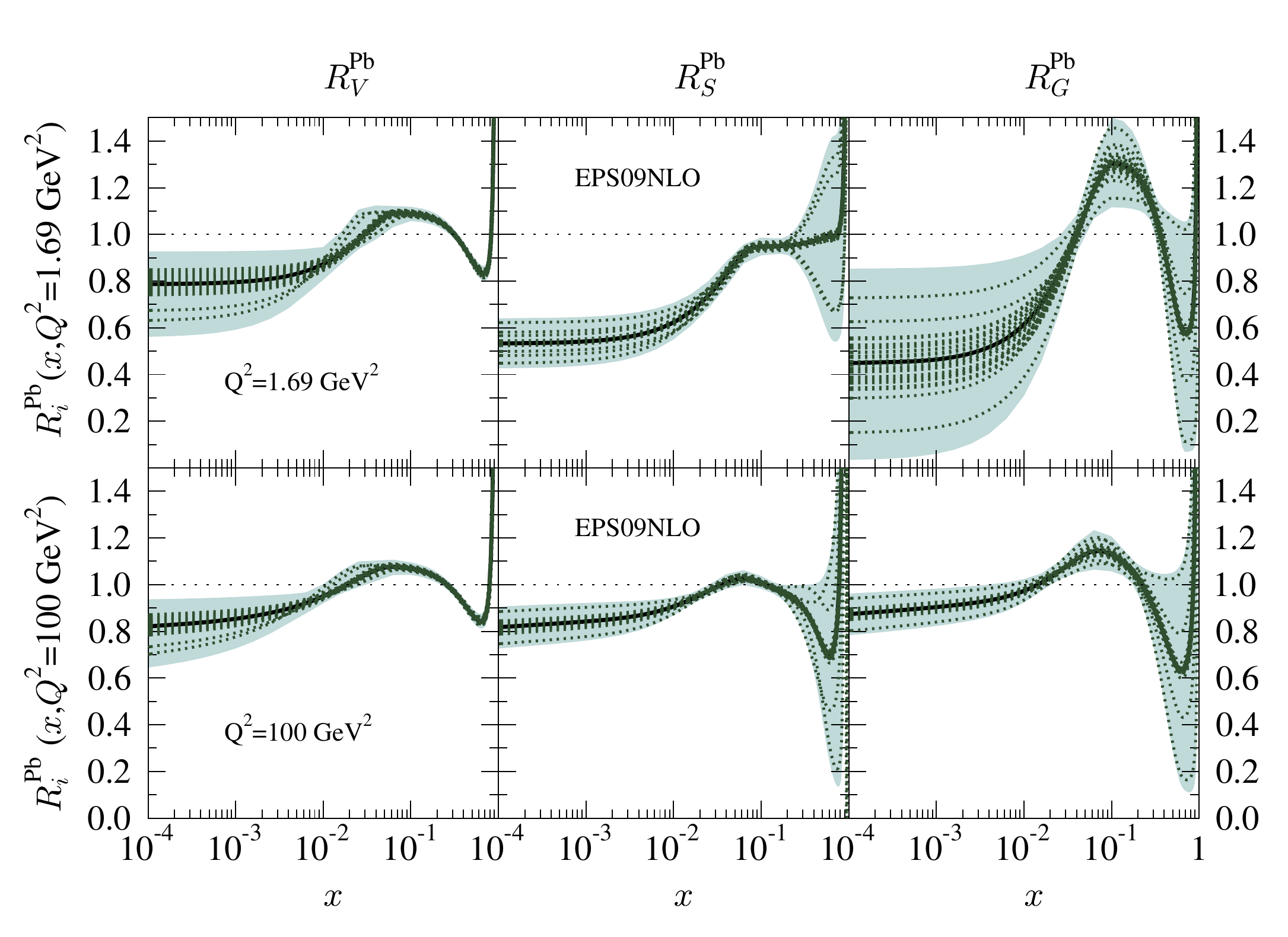} \includegraphics[scale=0.33]{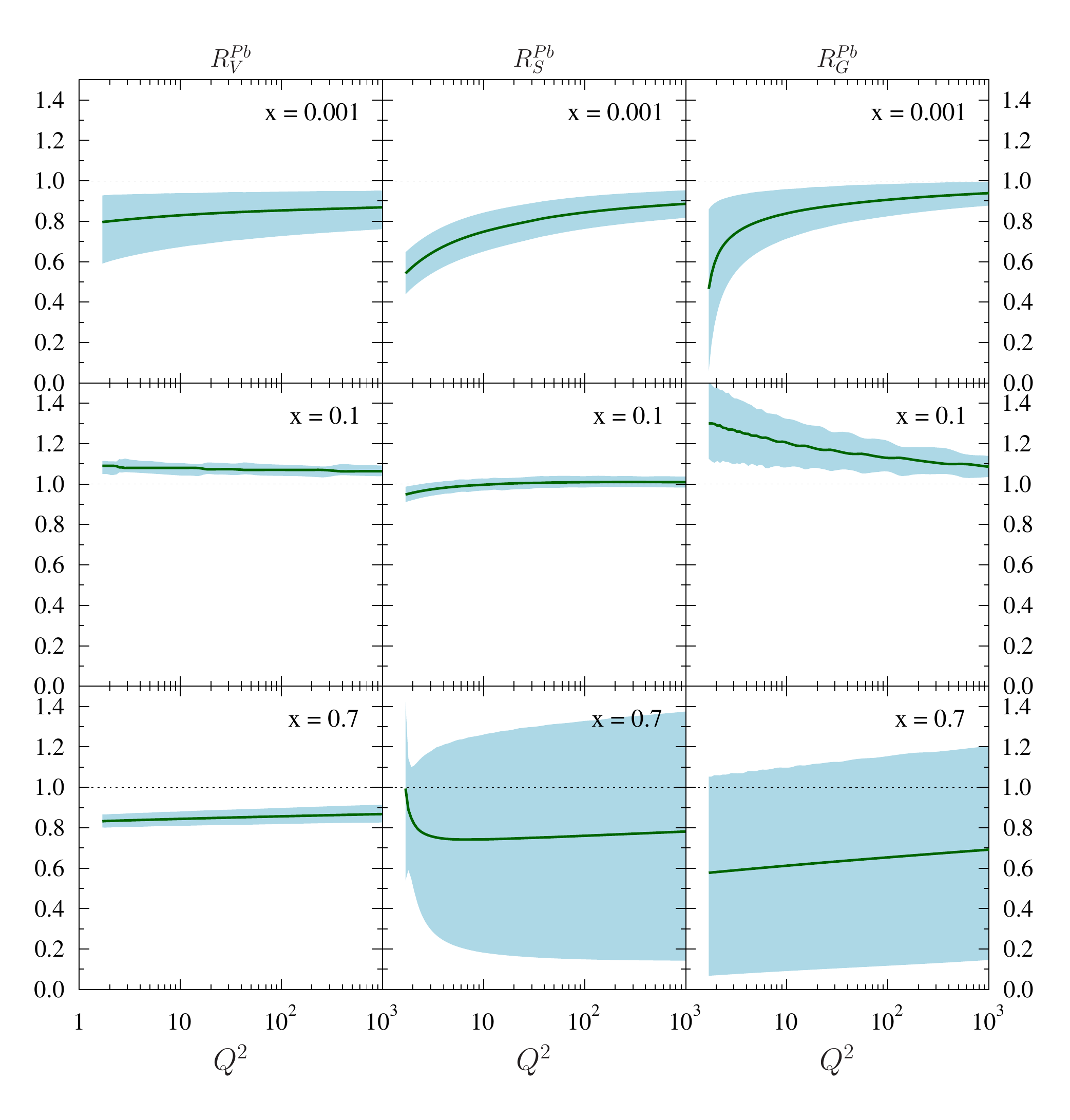}\hspace{-.7cm}
\vspace{-0.3cm}
\caption[]{ \textbf{Left:} The nuclear modifications $R_V^{\rm Pb}$, $R_S^{\rm Pb}$, $R_G^{\rm Pb}$ and their uncertainties at $Q^2=1.69 \, {\rm GeV}^2$ and 100 GeV$^2$. From \cite{EPS09}. \textbf{Right:} $Q^2$ evolution of these modifications at selected fixed values of $x$. }
\label{Fig:AllPDFs}
\end{figure}

\section{Results from EPS09}
\label{sec:EPS09_results}

Figure \ref{Fig:AllPDFs} (left) shows the average valence and sea quark and gluon modifications in a lead nucleus at the initial scale $Q_0^2$ and at a higher scale $Q^2=10$~GeV$^2$ according to the EPS09 NLO central set $S_0$ (solid lines) and the error sets $S_i^{\pm}$ (dotted lines). The shaded uncertainty band is computed using Eq.~(\ref{eq:X_Extremum3}) above. The largest uncertainties obviously reside at the smallest-$x$ and largest-$x$ gluons, since these are the gluon regions worst constrained by the data. In the DGLAP evolution, Fig.~\ref{Fig:AllPDFs} (right), the small-$x$ gluon uncertainties, however, quickly shrink, while the large-$x$ uncertainties not only remain but are also transferred into the large-$x$ sea quarks.

\begin{figure}[t!]
\center
\vspace{-0.5cm}
\hspace{-1.0cm}
\includegraphics[scale=0.35]{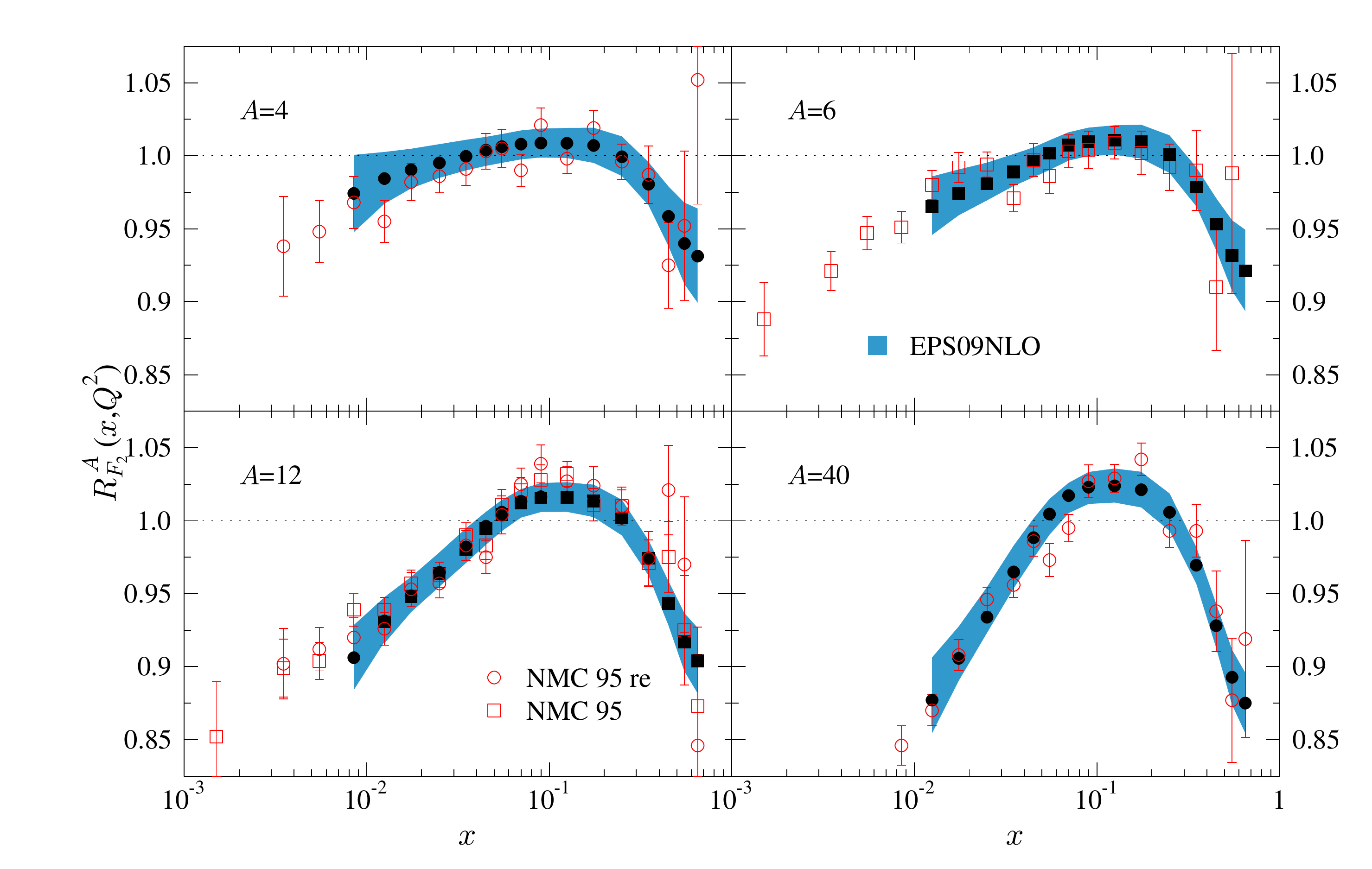}
\includegraphics[scale=0.33]{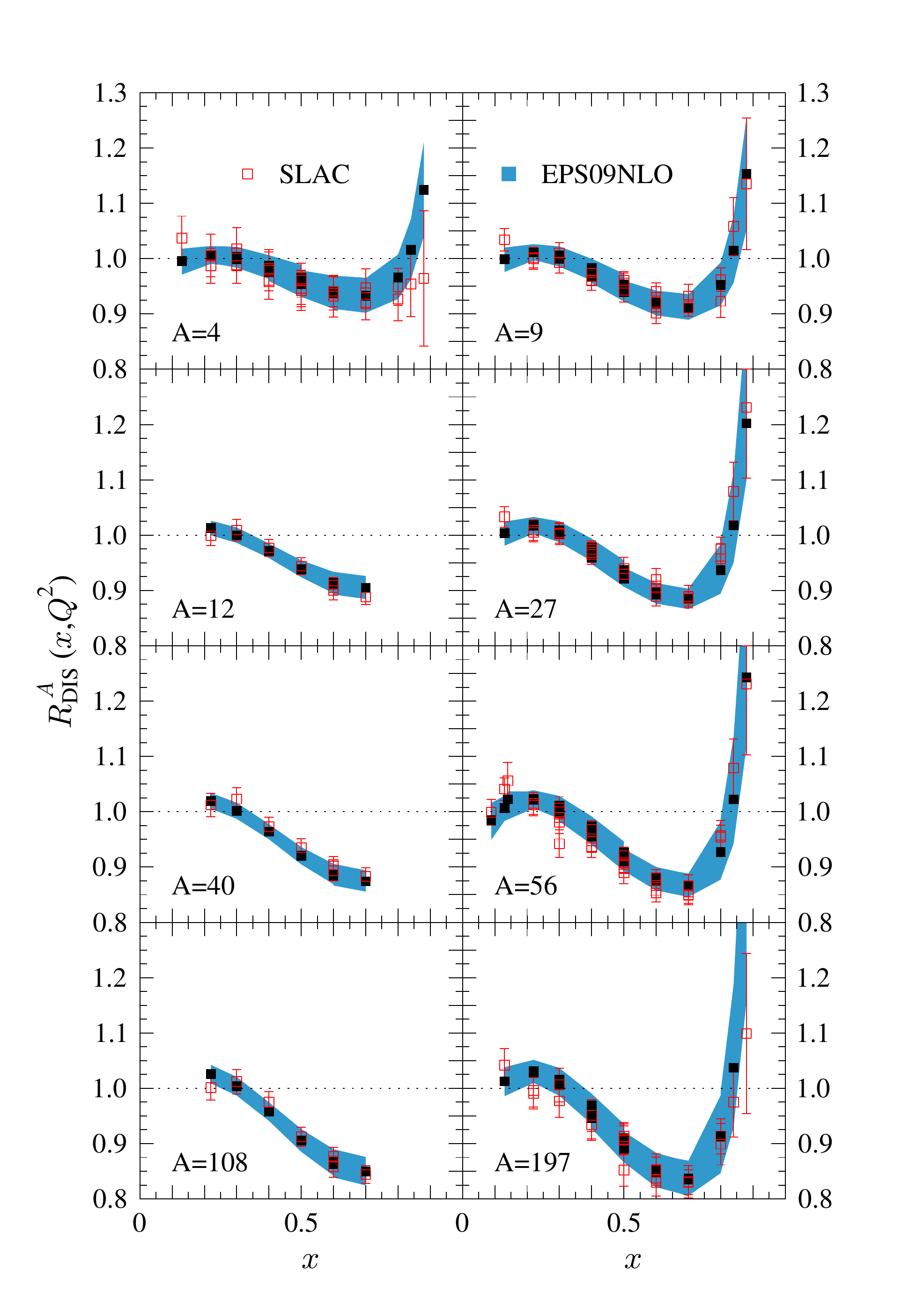}
\vspace{-0.3cm}
\caption[]{ \textbf{Left:} $R_{F_2}^A(x,Q^2)$ as computed in NLO using the EPS09 nPDFs (filled squares and error bands), and as measured by NMC 95 \cite{Arneodo:1995cs,Amaudruz:1995tq} (open symbols with error bars).
\textbf{Right:} The same for $R_{\rm DIS}^{\rm A}(x,Q^2)$ and SLAC data \cite{Gomez:1993ri}. From \cite{EPS09}.}
\label{Fig:RF2A1}
\end{figure}

Figure \ref{Fig:RF2A1} shows an example of the comparison of the EPS09 NLO results with the DIS data for the ratios
\begin{equation}
R_{F_2}^{\rm A}(x,Q^2) \equiv  \frac{F_2^A(x,Q^2)}{F_2^{\rm d}(x,Q^2)}, \quad
R_{\rm DIS}^{\rm A}(x,Q^2) \equiv  \frac{\frac{1}{A}d\sigma_{\rm DIS}^{l \rm
{A}}/dQ^2dx}{\frac{1}{2}d\sigma_{\rm DIS}^{l{\mathrm d}}/dQ^2dx}, 
\, {\rm with}\quad 
\sigma^{\ell + A \rightarrow \ell + X}_{\rm DIS} = \sum_{i=q,\overline{q},g} f_i^A(Q^2) \otimes \hat{\sigma}_{\rm DIS}^{\ell + i \rightarrow \ell + X}(Q^2). 
\end{equation}
From this figure, and from a similar comparison of the DY cross section ratios (see Fig.~4 in Ref.~\cite{EPS09}), we  confirm that with the choice $\Delta\chi^2=50$ the data uncertainties are transferred into the PDFs quite nicely: the error bands are indeed of the same size as the average error bars in the data. 

\begin{figure}[htb]
\center
\vspace{-0.5cm}
\includegraphics[scale=0.34]{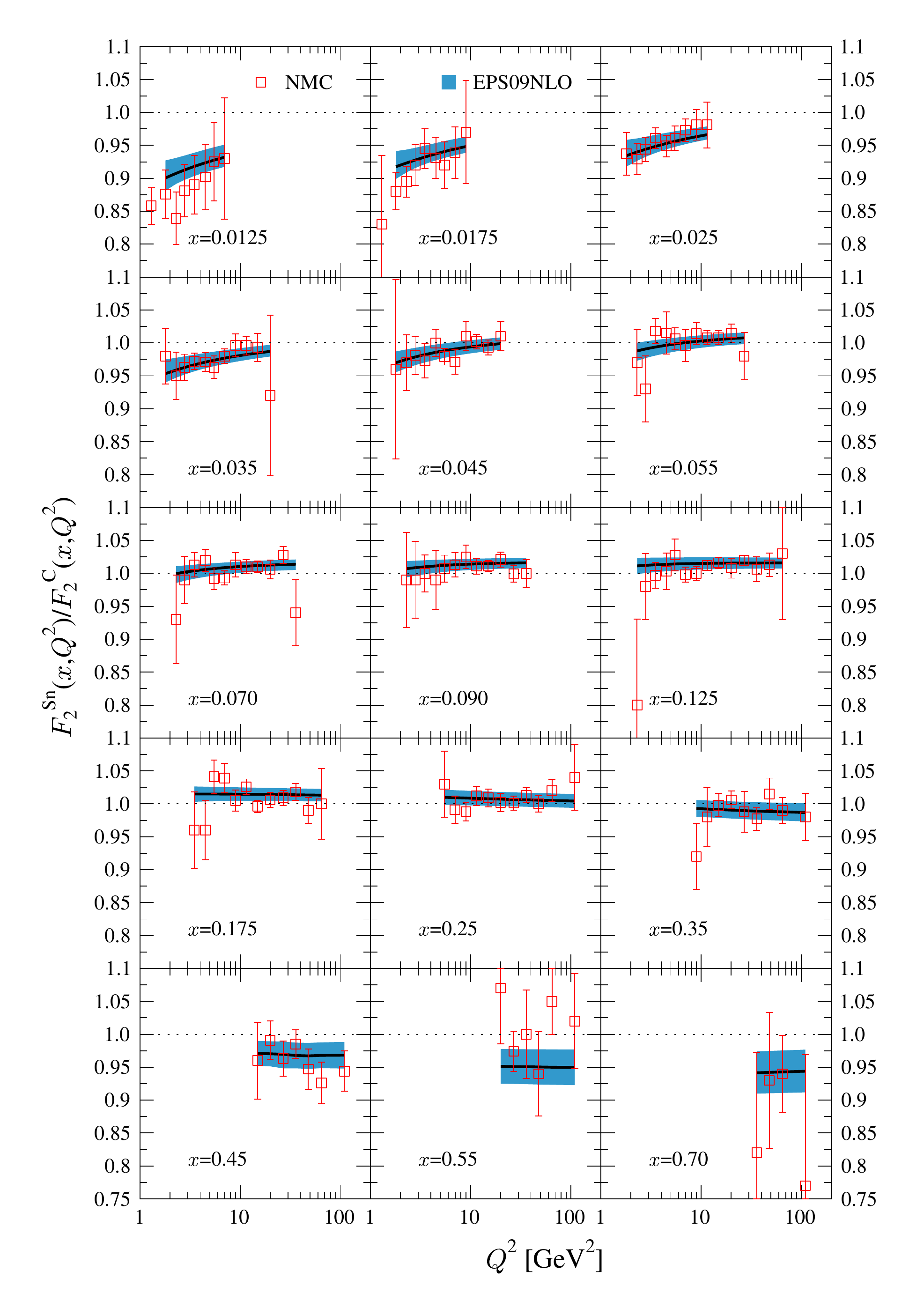}
\includegraphics[scale=0.415]{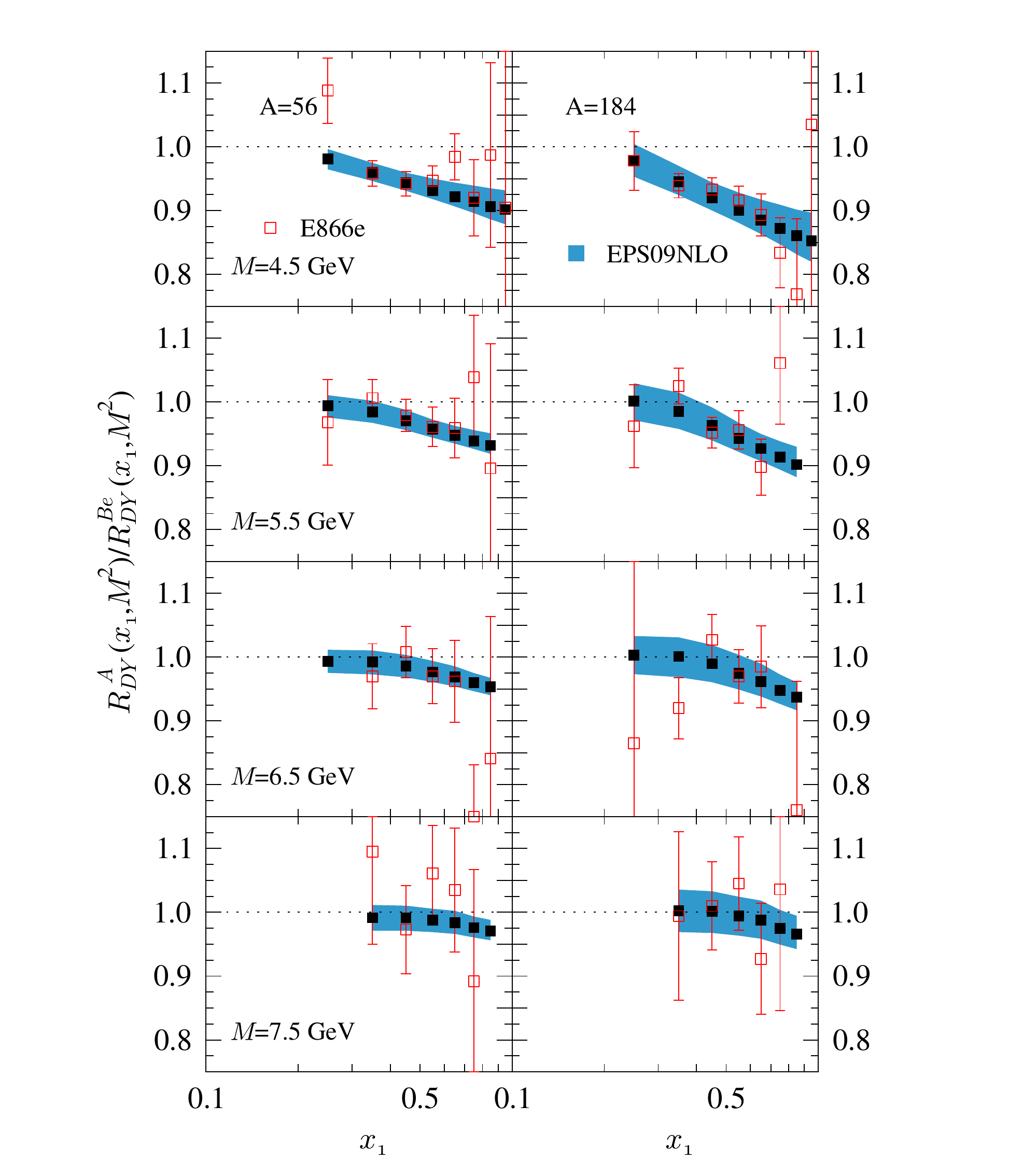}
\vspace{-0.3cm}
\caption[]{ \textbf{Left:} The calculated NLO scale evolution (solid lines and error bands) of 
  $F_2^{\mathrm{Sn}}/F_2^{\mathrm{C}}$ and the NMC data \cite{Arneodo:1996ru} for fixed values of $x$.
  \textbf{Right:} $R_{\rm DY}^{\rm A}(x_1,M^2)$ results from EPS09 NLO (filled squares and error bands) as a function of $x_1$ compared with the E866 \cite{Vasilev:1999fa} data (open squares). From \cite{EPS09}. }
\label{Fig:RF2SnC}
\end{figure}

The effects of the DGLAP evolution in the nuclear modifications are perhaps best illustrated by Fig.~\ref{Fig:RF2SnC} which shows the comparison of the EPS09 NLO results for $F_2^{\mathrm{Sn}}/F_2^{\mathrm{C}}$ (left) and for the DY cross section ratio (right) 
\begin{equation}
R_{\rm DY}^{\rm A}(x_{1},M^2) \equiv  \frac{\frac{1}{A}d\sigma^{\rm pA}_{\rm 
DY}/dM^2dx_{1}}{\frac{1}{2}d\sigma^{\rm pd}_{\rm DY}/dM^2dx_{1}},
\quad {\rm where} \quad
\sigma^{p + A \rightarrow l^+ l^- + X}_{\rm DY}  =  \sum_{i,j=q,\overline{q},g} f_i^p(M^2) \otimes f_j^A(M^2) \otimes \hat{\sigma}^{ij\rightarrow l^+ l^- + X}(M^2),
\end{equation}
and $x_1 = (M^2/\sqrt{s})e^y$, with $M^2$ being the invariant mass and $y$ the rapidity of the lepton pair.
The smallest-$x$ panel on the left is currently the directmost constraint for nuclear gluons at $x\sim{\cal O}(0.01)$.
Smaller-$x$ DIS data at perturbative scales would be badly needed to further constrain the nuclear gluon distributions at smaller $x$.

\begin{figure}[!h]
\vspace{-1.0cm}
\center
\hspace{-.5cm}\includegraphics[width=20pc]{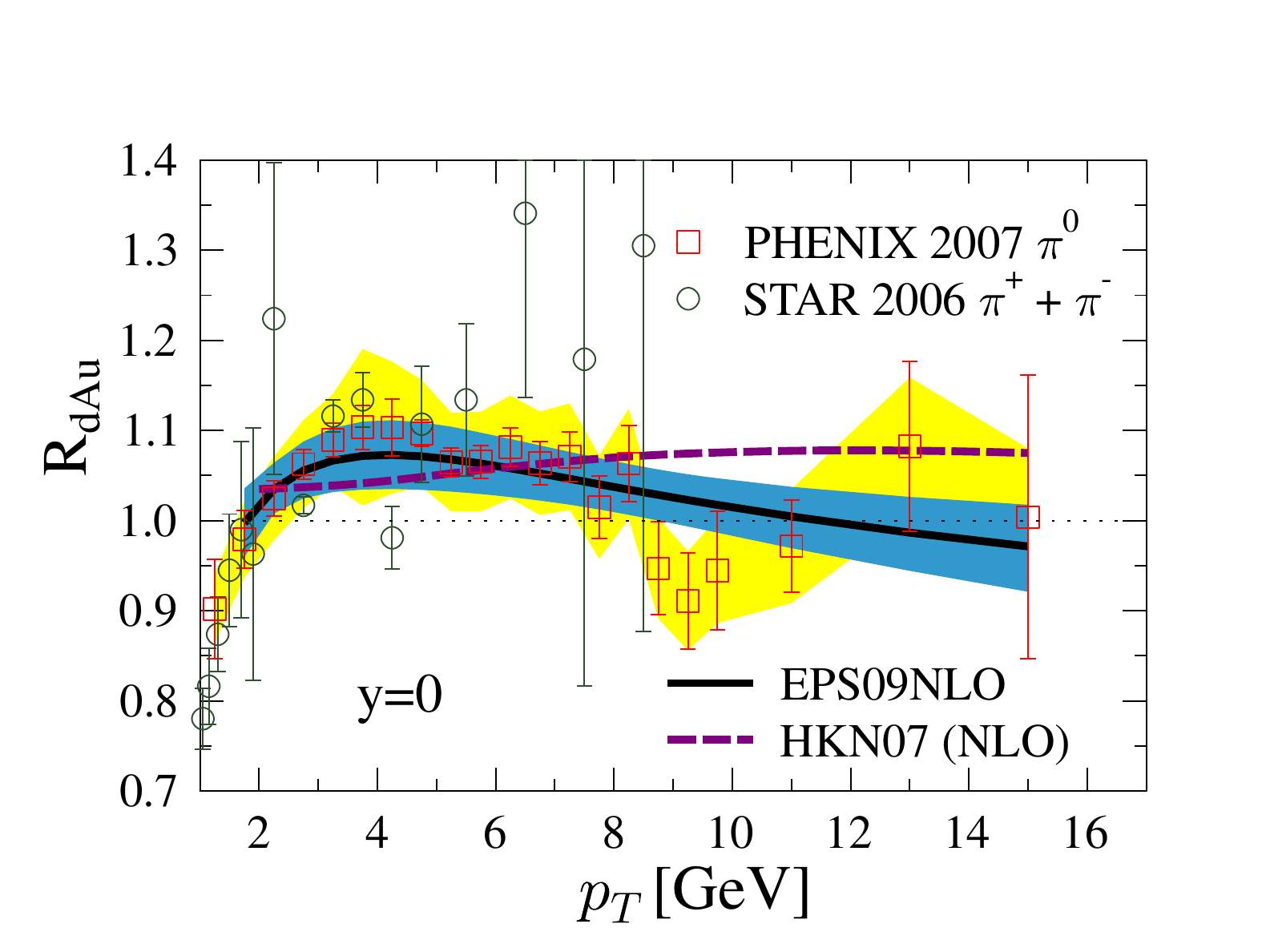}
\hspace{-.0cm}\includegraphics[scale=0.38]{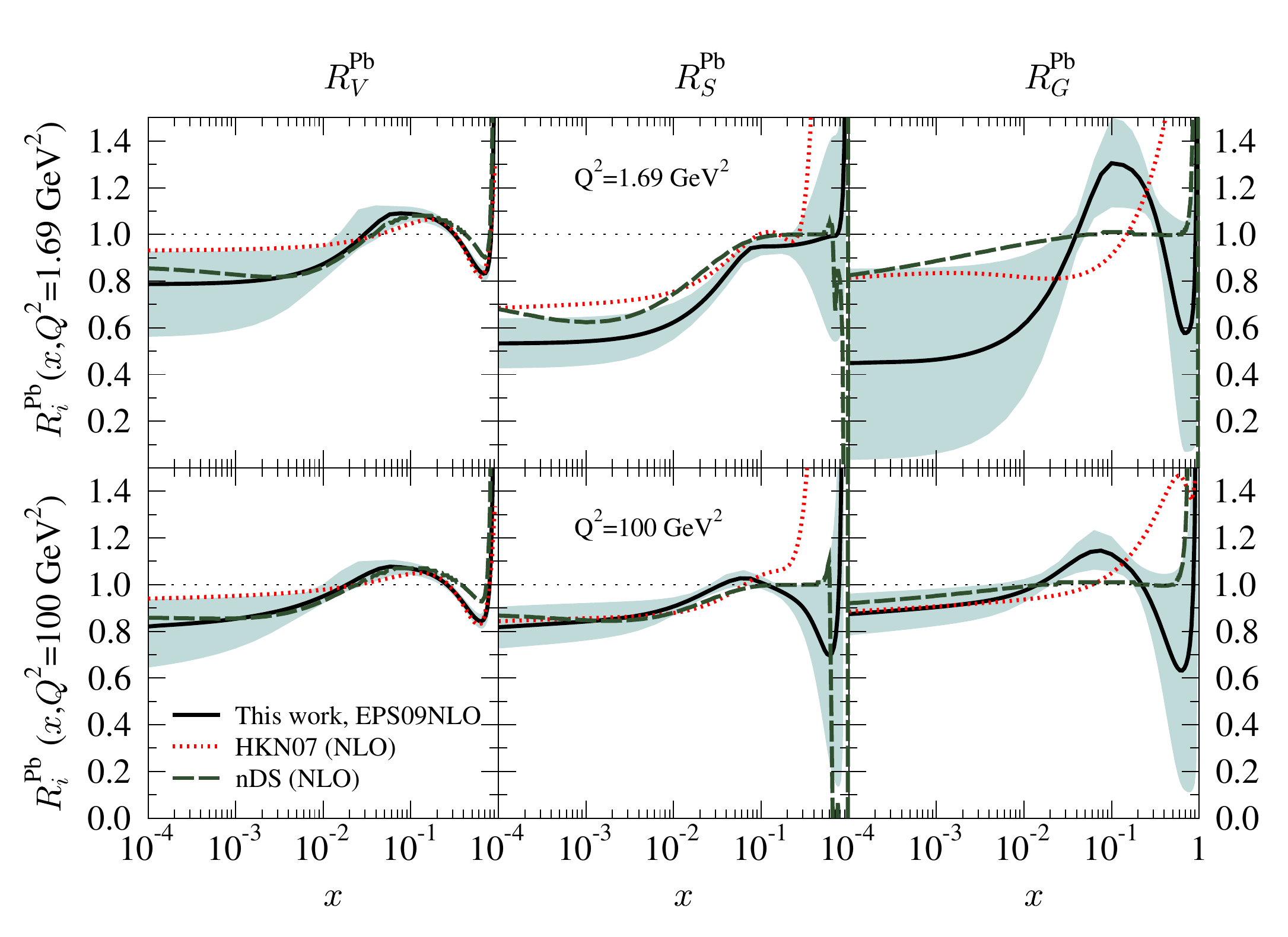}
\vspace{-0.5cm}
\caption[]{ 
\textbf{Left:} {$R_{\rm dAu}$ from EPS09 NLO (solid line, blue error band) at $y=0$ for inclusive $\pi^0$ production and the PHENIX \cite{Adler:2006wg} data (open squares). The error bars show the statistical and the yellow band the systematic errors. The prediction from HKN07 (NLO) is shown by the dashed line. The data have been multiplied by $f_N = 1.03$, see Eq.~(\ref{eq:chi2}), which is an output of the EPS09 analysis. The STAR data \cite{Adams:2006nd} (open circles), multiplied by $f_N = 0.90$, are shown only for comparison.}
\textbf{Right:} The average valence and sea quark, and gluon modifications at $Q^2 = 1.69 \, {\rm GeV}^2$ and $Q^2 = 100 \, {\rm GeV}^2$ for Pb nucleus from the NLO HKN07~\cite{Hirai:2007sx}, nDS~\cite{deFlorian:2003qf} and EPS09. From \cite{EPS09}.}
\label{Fig:PHENIX}
\end{figure}

Comparison of the EPS09 NLO results for the inclusive mid-rapidity pion production ratio in minimum-bias d+Au and  p+p collisions at RHIC,
\begin{equation}
R_{\rm dAu}^{\pi} \equiv \frac{1}{\langle N_{\rm coll}\rangle} \frac{d^2 N_{\pi}^{\rm dAu}/dp_T dy}{d^2 N_{\pi}^{\rm pp}/dp_T dy} \stackrel{\rm min. bias}{=} \frac{\frac{1}{2A} d^2\sigma_{\pi}^{\rm dAu}/dp_T dy}{d^2\sigma_{\pi}^{\rm pp}/dp_T dy},
\end{equation}
where
\begin{equation}
\sigma^{A+B \rightarrow \pi + X}\hspace{-0.2cm} = \hspace{-0.4cm}\sum_{i,j,k=q,\overline{q},g} f_i^A(p_T^2) \otimes f_j^{B}(p_T^2) \otimes \hat{\sigma}^{ij\rightarrow k + X}(p_T^2) \otimes D_{k \rightarrow \pi}(p_T^2), 
\end{equation}
with the PHENIX data is shown in Fig.~\ref{Fig:PHENIX} (left).
We use the KKP fragmentation functions \cite{Kniehl:2000fe} here, but we have checked that with the newer ones, AKK08 \cite{Albino:2008fy} or nDSS \cite{de Florian:2007hc,deFlorian:2007aj}, the results would be the same. The NLO code by Aversa et al. \cite{Aversa:1988vb} from the {\ttfamily INCNLO} \cite{INCNLO} compilation has been utilized after changing the order of integrations to speed up the computation (see \cite{EPS09}). Interestingly, and importantly, a good and essentially tensionless fit of these RHIC data is obtained simultaneously with the DIS and DY data. This data set, especially the downward trend at $p_T>5$~GeV, also offers a valuable constraint for the antishadowing and EMC effects in the gluon distributions. 

Figure~\ref{Fig:PHENIX} also compares the NLO results from the global DGLAP analyses HKN07~\cite{Hirai:2007sx}, nDS~\cite{deFlorian:2003qf} and EPS09. The main difference between these sets is the existence of the gluon antishadowing/EMC-effect structure present only in EPS09. This is mainly due to the fact that the PHENIX $\pi^0$ data, which shows such behaviour, are included only in EPS09. The number of data points included in the EPS09/HKN07/nDS analyses is, correspondingly, 929/1241/420, while the obtained $\chi^2$ per data point is, correspondingly, 0.79/1.2/0.72.

\begin{figure}[ht!]
\center
\vspace{-0.5cm}
\includegraphics[scale=0.40]{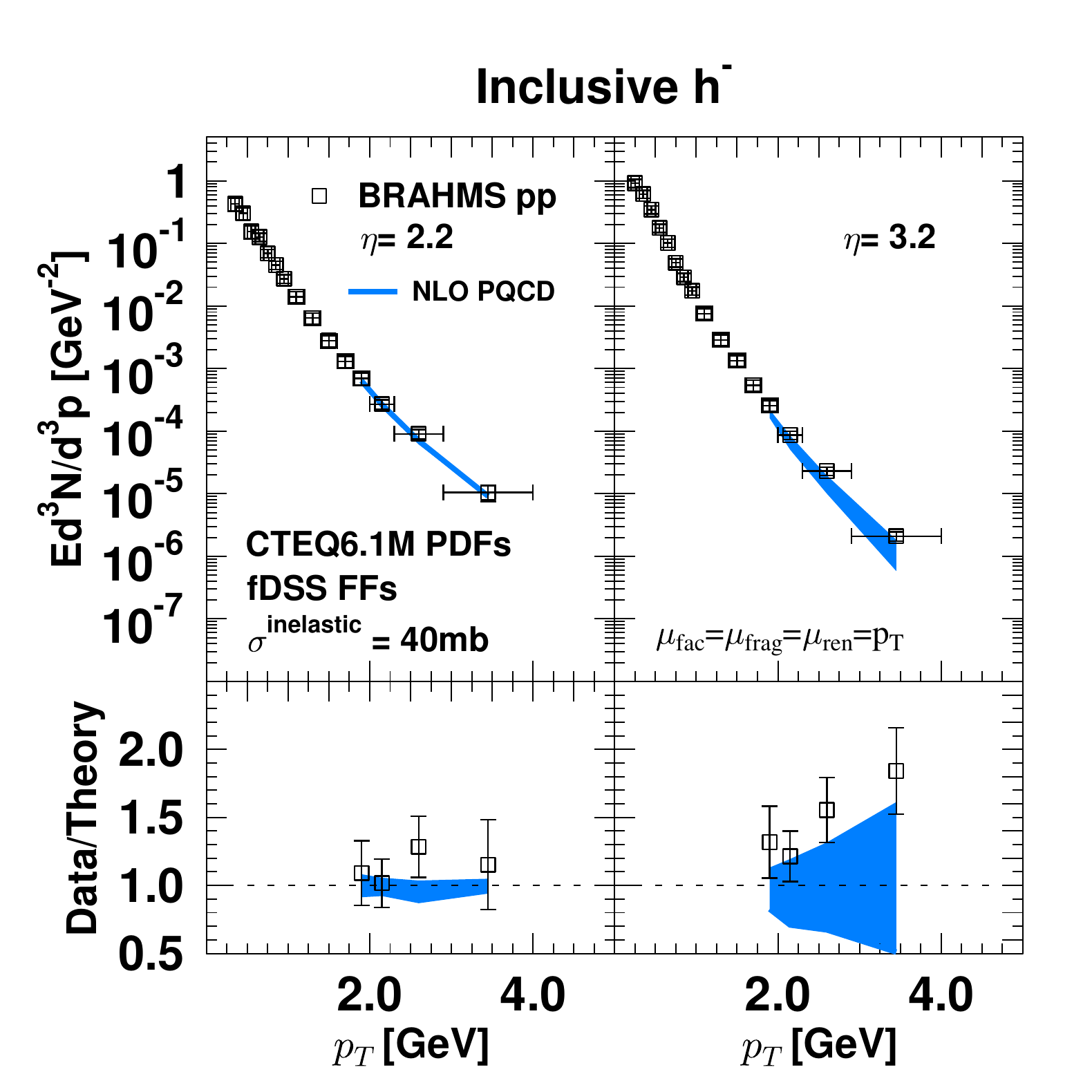}
\includegraphics[scale=0.40]{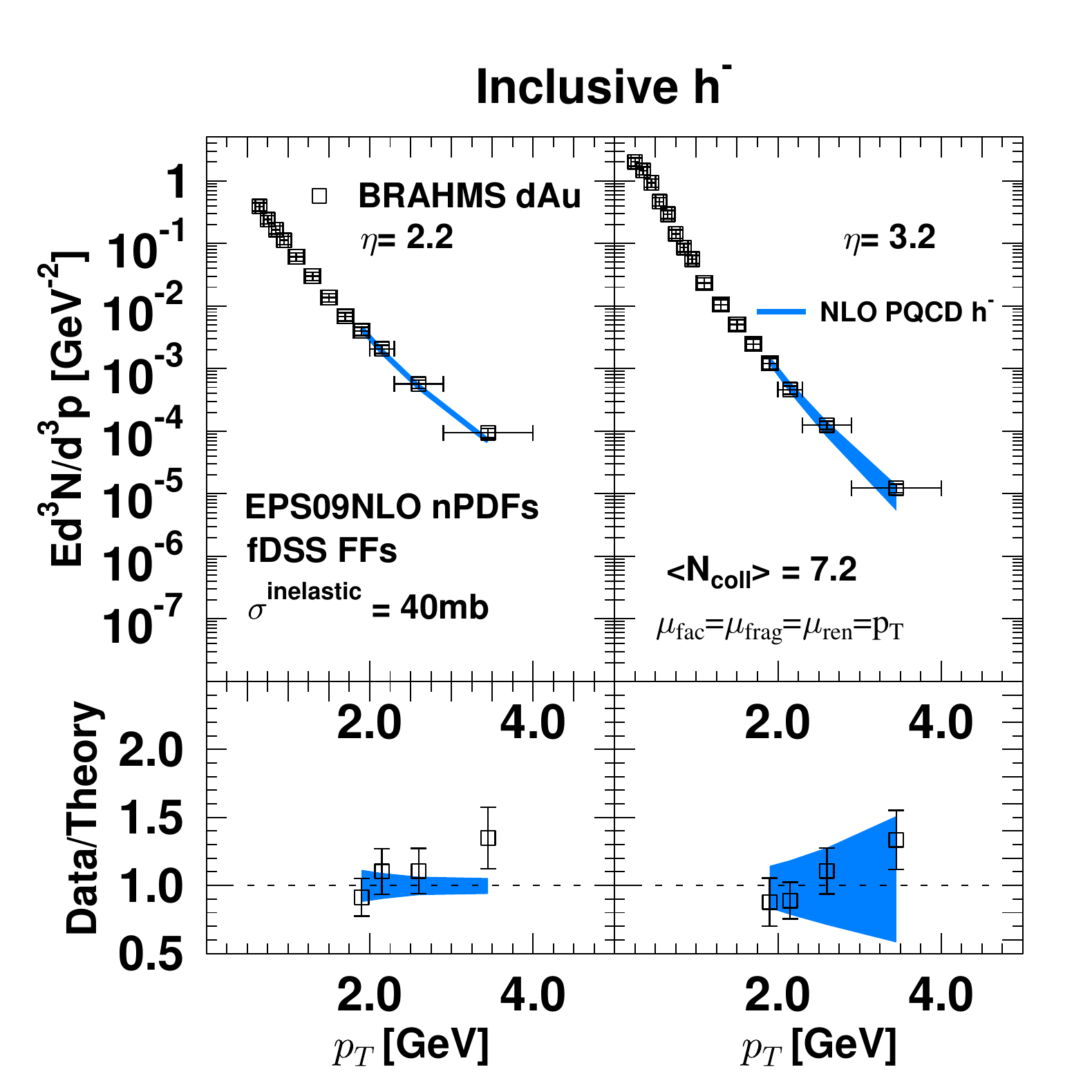}
\vspace{-0.3cm}
\includegraphics[scale=0.4]{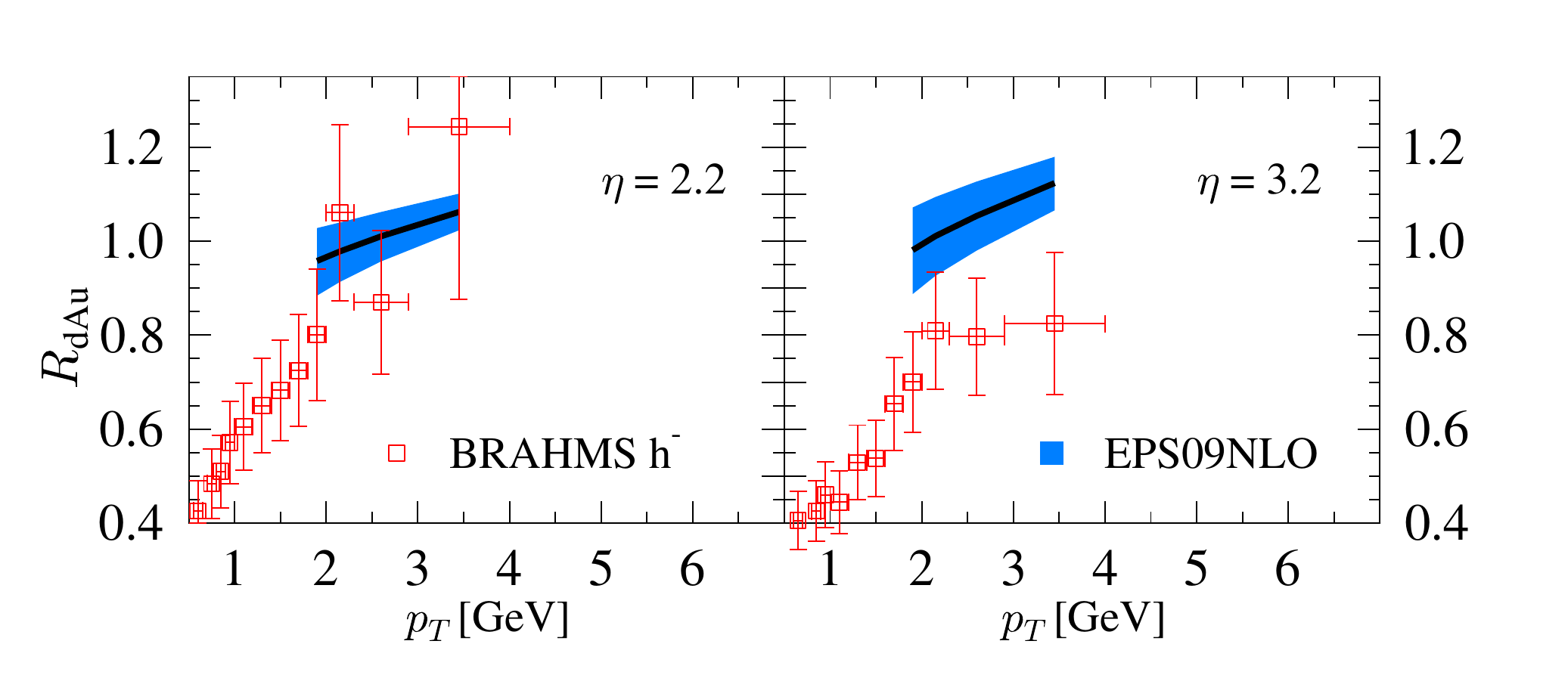}
\caption[]{ 
\textbf{Top left:} $EdN/d^3p$ of negatively charged hadrons in p+p and d+Au collisions vs. $p_T$. The data (open squares) with error bars are from \cite{Arsene:2004ux} with statistical and systematical errors added in quadrature. The error band is the free-proton PDF uncertainty given by CTEQ6.1M. 
\textbf{Top right:} The same for d+Au, except now the error bands are obtained using the error sets of both CTEQ6.1M and EPS09.
\textbf{Bottom:} The corresponding ratio $R_{\rm dAu}$. The error bands here correspond to the EPS09NLO error sets.
From \cite{EPS09}}
\label{Fig:BRAHMS1}
\end{figure}

\section{EPS09 vs. BRAHMS forward-rapidity data}
\label{sec:EPS09_vs_BRAHMS}
As an example of using the EPS09 package, we consider the forward-rapidity hadron production,  measurements for which have been done by BRAHMS  \cite{Arsene:2004ux}. Figure~\ref{Fig:BRAHMS1} shows the invariant distributions $EdN/d^3p$ in p+p (left) and in d+Au collisions (right). The error bands shown in the p+p case have been obtained by using the central set together with the 40 error sets from CTEQ6.1M, and the ones for the d+Au case by using the central sets from CTEQ6.1M and EPS09NLO together with their 40(CTEQ6.1M) + 30 (EPS09NLO) error sets as explained in Sec.~4 of Ref.~\cite{EPS09}. We notice that the data are systematically above the NLO pQCD calculation in the p+p case at $\eta=3.2$, while in both nuclear cases considered, and also for $\eta=2.2$ in p+p, there is an agreement. This discrepancy is why these BRAHMS data was excluded from the EPS09 analysis. After this, it is not a surprise that the data for the nuclear ratio $R_{\rm dAu}$ at $\eta=3.2$ and $p_T>2$~GeV, as shown by the bottom panel of Fig.~\ref{Fig:BRAHMS1}, lie clearly below the EPS09NLO prediction. Thus, the NLO EPS09 global analysis suggests that the strong suppression seen in the BRAHMS $R_{\rm dAu}$ at $\eta=3.2$ and $p_T>2$~GeV is due to an excess in the p+p data, not due to a suppression in the d+Au data.

\section{Summary and outlook}
\label{sec:summary}

The conclusion from the EPS09 NLO global DGLAP analysis is that there is an excellent agreement between NLO pQCD and the hard-process nuclear data for deeply inelastic lepton-nucleus scattering, DY-dilepton production in p+$A$, and central-rapidity $\pi^0$ production in d+Au collisions in the nPDFs' kinematical range 
$0.005 \lesssim x \le 1$, $1.69 \le Q^2 \lesssim 150$~GeV$^2$. The goodness of the global fit with 929 data points is
$\chi^2/N$ = 0.79, and remarkably, no significant tension between the data sets from different types of processes is found.
Collinear factorization seems to work well in describing high-energy inclusive nuclear hard processes.

The EPS09 nPDF package contains the central set (best fit) + 30 error sets, both in NLO and LO, and they are downloadable at https://www.jyu.fi/fysiikka/en/research/highenergy/urhic/nPDFs. 
With these error sets, it is now, finally, possible for anybody to study how the nPDF-uncertainties propagate into nuclear hard cross sections. This should be particularly useful for the precision analysis of hard probes as QCD matter signatures, and also, e.g., for planning detector upgrades in the near future.

Further tests of factorization and nPDFs will be provided e.g. by RHIC data for direct photons in d+Au, and both RHIC and LHC data in Au+Au collisions. The corresponding pQCD cross sections have been discussed e.g. in Ref.~\cite{Arleo:2007js}, and also EPS09NLO-related direct photon studies both for RHIC and LHC have been launched \cite{Arleo_Eskola}. To constrain the nuclear quark distributions better, also (anti)neutrino-nucleus DIS data should be added into the global analysis of the nPDFs. Recent studies \cite{Paukkunen:2010hb}, suggest that the agreement with the CHORUS, CDSHSW and also most of the NuTeV data (see \cite{Paukkunen:2010hb} for references) is in fact quite good. Before including the neutrino-DIS data into the EPS09-type global analysis consistently, however, one needs to extend the analysis to a general-mass framework. 

Further tests of factorization will also be given by heavy-quark production and forward-rapidity pion data in d+Au collisions at RHIC. To properly test the applicability of factorization at the available highest energies as well -- and to get the most reliable comparison baselines for the nuclear hard processes -- it would quite obviously be very important to have the p+Pb runs scheduled at the LHC, too. For recent studies of constraining the nPDFs at the LHC, see e.g. Refs.~\cite{Paukkunen:2010qg,QuirogaArias:2010wh}. The most direct constraints for the still unconstrained lowest-$x$ nuclear gluons one would, however, get from the planned future colliders eRHIC and in particular LHeC. 

Finally, regarding a longer-term outlook for the global analyses,  in order to obtain a minimal number of uncorrelated error sets simultaneously for the free-proton PDFs and for the nPDFs, one should eventually combine the global analyses for the free and bound proton PDFs into one single master analysis. 
\vspace{0.1cm}

\noindent \textbf{Acknowledgements:} KJE thanks the Vilho, Yrj\"o and Kalle V\"ais\"al\"a fund of the Finnish Academy of Science and Letters, and the Academy of Finland (Project nr. 133005) for financial support.
\vspace{-0.3cm}







\end{document}